\documentclass[12pt,ams,a4paper,nofootinbib]{revtex4}
\usepackage{amsmath}
\usepackage{graphicx}
\usepackage[usenames,dvipsnames]{color}

\newcommand{\re}{\ref}
\newcommand{\be}{\begin{equation}}
\newcommand{\ee}{\end{equation}}

\newcommand{\la}{\label}
\newcommand{\ber}{\begin{eqnarray}}
\newcommand{\eer}{\end{eqnarray}}

\begin{document}

\title{On calculating response functions via their Lorentz integral transforms}

\author{  Victor D. Efros$^{1,2}$, Winfried Leidemann$^{3,4}$, and Veronika Yu. Shalamova$^{1}$
  }

\affiliation{$^{1}$National Research Centre "Kurchatov Institute", 123182 Moscow, Russia\\
$^{2}$National Research Nuclear University MEPhI(Moscow Engineering Physics Institute)\\
$^{3}$Dipartimento di Fisica, Universit\`a di Trento, I-38123 Trento, Italy\\
$^{4}$INFN-TIFPA Trento Institute of Fundamental Physics and Applications, I-38123 Trento, Italy
}
 
\date{\today}
 
\begin{abstract}
The accuracy of reconstruction of a response function from its Lorentz integral transform is studied in an exactly solvable model. 
An inversion procedure is elaborated in detail and features of the procedure are studied.
Unlike  
results in the literature pertaining to the same model, the response function is reconstructed from its Lorentz 
integral transform 
with rather
high accuracy.
\end{abstract}

\bigskip

\maketitle
\section{introduction}

We address the issue of computing the
response functions 
\be R(E)=\sum_n|(\Psi_n,O\Psi_0)|^2\delta(E-E_n)+
\sum\!\!\!\!\!\!\!\int df|(\Psi_f,O\Psi_0)|^2\delta(E-E_f)\la{1}\ee
of quantum mechanical systems.
Here $\Psi_n$ and $\Psi_f$ represent a complete set 
of bound plus
continuum--spectrum states of the Hamiltonian of a problem, $\Psi_0$ is the initial state, and  $O$ is a transition operator.
The subscript~$f$ denotes collectively a set of continuous and discrete 
variables labeling a state which is symbolized by the summation over integration notation.
The states are orthonormalized, \mbox{$(\Psi_n,\Psi_{n'})=\delta_{n,n'}$} and
\mbox{$(\Psi_f,\Psi_{f'})=\delta(f-f')$}. 
Eq.~(\re{1})  represents the response of a system to an external probe which is an important observable
quantity.

When the number of particles in a system exceeds two or three
it is not possible in practice to compute the responses  directly from their definition~(\re{1}).
But they can be reconstructed from their integral transforms. 
In particular, the Lorentz integral transform (LIT)~\cite{ELO94} is efficient to this aim
when  expansions over many--body basis functions are used in order to solve the arising bound--state like problem. 
Note also that while the quantity~(\re{1}) is an inclusive one, an ability to calculate quantities of a similar structure
makes possible obtaining exclusive amplitudes of general--type multichannel reactions, see the review~\cite{rev} and references 
therein. 

In Ref.~\cite{suz} an attempt to verify the LIT approach has been undertaken employing a model
for the three--particle
photodisintegration. The model involves one degree of freedom and therefore
can be solved exactly. The calculations of Ref.~\cite{suz}  have led to the results
 which are at variance with the exact solution
and have nonphysical features.
In view of this, we reconsider the matter in the present paper. Besides, we discuss in detail features
of the inversion of the integral transform. Such a discussion is useful to perform many--body calculations and it was not
presented in previous work on the subject.

Various aspects of calculating and inverting LITs have been considered in 
Refs.~\cite{ELO94,ELO99,efr99,reis,nir,leid1,leid2,leid3,schw,efr,andr}.  
Of them, Refs.~\cite{ELO94,ELO99,efr99,reis,nir,leid1,leid2,leid3} deal with the 
"standard" inversion method,
in Refs.~\cite{schw,efr} other inversion methods are tried and/or constructed, 
and in Ref.~\cite{andr} both the standard and other methods are studied. 
As in Ref.~\cite{suz}, 
 the standard inversion method is employed in the present work. 

\section{Formulation of the problem}

In the model of Ref.~\cite{suz} the dipole
photodisintegration of the bound 
state of three particles interacting  via a hypercentral potential is considered.
Up to an energy--independent constant, the model
 is equivalent to a one--body problem in which the hypercentral potential is represented as
a central one and a nucleon with the "orbital momentum" 3/2 bound in this  
potential 
passes to the  continuum state with  the "orbital momentum" 5/2.

Denote the potential as~$V(\rho)$. The initial state~$\chi_0(\rho)$ is determined from the equation
\be
-\frac{\hbar^2}{2m}\frac{d^2\chi_0(\rho)}{d\rho^2}+\left[\frac{\hbar^2}{2m}\frac{l_0(l_0+1)}{\rho^2}+V(\rho)
-E_0\right]\chi_0(\rho)=0
\la{11}\ee
with $l_0=3/2$.  
The normalization is 
\[\int_0^\infty d\rho\, \chi_0^2(\rho)=1.\]

The final state ~$\chi_E(\rho)$ is determined from the equation
\be
-\frac{\hbar^2}{2m}\frac{d^2\chi_E(\rho)}{d\rho^2}+\left[\frac{\hbar^2}{2m}\frac{l_1(l_1+1)}{\rho^2}+V(\rho)
-E\right]\chi_E(\rho)=0
\la{22}\ee
with $l_1=5/2$.   
Beyond the range of the potential $\chi_E(\rho)$ behaves as
\be\sqrt{m\rho/\hbar^2}\left[ J_{l_1+1/2}(k\rho)\cos\delta-N_{l_1+1/2}(k\rho)\sin\delta\right]\la{33}\ee
where $(\hbar k)^2/(2m)=E$. The continuum wave functions are normalized as follows,
\be
\int_0^\infty d\rho\,\chi_E(\rho)\chi_{E'}(\rho)=\delta(E-E').\la{de}
\ee
We shall calculate the response function given by the expression
\be 
r(E)= \left[\int_0^\infty d\rho\,\chi_E(\rho)\rho\chi_0(\rho)\right]^2.\la{5}
\ee
At $df=dE_f$, which is our case, Eq.~(\re{1}) with $O$ being the dipole operator
turns to such an expression up to an energy independent constant.
The expression (\re{5}) differs by such a constant from that adopted for $R(E)$ in Ref. \cite{suz}.


As said above, in the many--body case  
the method of integral transforms is employed to compute the response functions. As in Ref.~\cite{suz}, as
a test of the ability of the LIT approach we 
shall calculate the response~(\re{5}) both directly and via its LIT and we shall compare the results.

To perform the latter of the two mentioned calculations we employ the solution
to the inhomogenious equation
\be
-\frac{\hbar^2}{2m}\frac{d^2\chi_\sigma(\rho)}{d\rho^2}+\left[\frac{\hbar^2}{2m}\frac{l_1(l_1+1)}{\rho^2}+V(\rho)
-\sigma\right]\chi_\sigma(\rho)=\rho\chi_0(\rho).
\la{6}\ee
Its right--hand side includes the initial--state wave function from Eq.~(\re{11}), 
and
$\sigma$ is a complex energy.
At large $\rho$ values the solution $\chi_\sigma(\rho)$ tends to zero decreasing exponentially.
Let us use the notation 
\be \Phi(\sigma_R,\sigma_I)=\int_0^\infty d\rho\,\chi_\sigma^*(\rho)\chi_\sigma(\rho),\la{phi}\ee
where $\sigma_R$ and $\sigma_I$ denote the real and the imaginary part of $\sigma$.
The response (\re{5}) can be found~\cite{ELO94} as the solution to the integral equation
\be
\Phi(\sigma_R,\sigma_I)=\int_0^\infty dE\,\frac{r(E)}{(E-\sigma_R)^2+\sigma_I^2}.\la{ieq}
\ee

\section{Solving the dynamics equations}

We adopt the same potential, $V(\rho)=V_0\exp(-\kappa\rho^2)$, $V_0=-75$~Mev, and $\kappa=0.16$~fm$^{-2}$, 
and the same value of $\hbar^2 /m= 41.47106$~MeV~fm$^2$ as in Ref.~\cite{suz}.
We solve the above equations (\re{11}), (\re{22}), and (\re{6}) employing expansions  over the radial oscillator functions
$\phi_n(\rho)$,
\be
\phi_n(\rho)={\cal N}_{nl}\rho_0^{-1/2}
x^{l+1}
L_n^{l+1/2}(x^2)
e^{-x^2/2},\qquad \int_0^\infty d\rho\, \phi_n(\rho)\phi_m(\rho)=\delta_{mn}.
\la{osc}\ee
Here $l$ equals  either 3/2 or 5/2,  ${\cal N}_{nl}$ is the normalization constant, and $x=\rho/\rho_0$. 
The oscillator radius~$\rho_0$ has been chosen to be 2.0~fm in all the calculations.

In the oscillator representation (\re{osc}) with $N_{max}$ basis functions retained
the bound--state problem (\re{11}) turns to the algebraic eigenvalue problem
$(H-E_0)\chi=0$ where $H$ is the \mbox{$N_{max}$--size} matrix corresponding to the operator
 from Eq.~(\re{11}) and $\chi$ is the 
column
of the expansion coefficients. The $H$ matrix has been calculated analytically.
The problem was solved with the method of inverse iteration, 
\mbox{$(H-E_{tr})\chi^{(n)}={\bar \chi}^{(n-1)}$}, where $E_{tr}$ is an energy sufficiently close to the $E_0$ eigenvalue,
$n=1,2,\ldots$,
and ${\bar \chi}^{(n)}$ denotes a $\chi^{(n)}$ column renormalized to unity. The  $\chi^{(0)}$ column may be chosen
arbitrarily provided it is not orthogonal to the solution sought for. We chose it to be  \mbox{$(\chi^{(0)})_n=\delta_{1n}$} and we
chose $E_{tr}=-3.5$~MeV. The iteration process terminated when the norm 
\mbox{$||{\bar \chi}^{(n)}-{\bar \chi}^{(n-1)}||$} became
smaller then $10^{-14}$. The required numbers of iterations equaled six or seven. The sets of linear equations here
and in all the cases below were solved with the help of the $LU$ decomposition of the matrices, see e.g. \cite{rec}.

In Table 1 the trend of convergence, 
of the energy $E_0$ and radius $\langle \rho^2\rangle^{1/2}$ of the bound state obtained is shown.
The quantity
$N_{max}$ denotes the number of the  oscillator functions~(\re{osc}) retained in the calculation.

\begin{table}[h]
\caption{Dependence of the energy $E_0$ [MeV] and the radius $\langle \rho^2\rangle^{1/2}$ [fm] of the bound state on the
number $N_{max}$ of the functions (\re{osc}) retained in the calculation.}
\vspace{0.2cm}
\begin{tabular}{|c|c|c|}
\hline
$N_{max}$& $E_0$ & $\langle  \rho^2\rangle^{1/2}$\\
\hline
25 & -3.492627476426842  & 3.18321989816014\\
\hline
150& -3.492628451703556 & 3.18326851163143  \\
\hline
300& -3.492628451703560  & 3.18326851163147 \\
\hline
\end{tabular}
\end{table}
In the $N_{max}=300$ case the calculations were done with the quadrupole 
precision. The values of $\langle \rho^2\rangle^{1/2}$ listed in Table I of Ref. \cite{suz} are  not correct.

The integral transforms $\Phi(\sigma_R,\sigma_I)$
 were calculated from Eq.~(\re{phi}) as functions of $\sigma_R$  at fixed values of $\sigma_I$.
As an additional test, the sum rule for the transform has been calculated. One has  \cite{schw}
\be (\sigma_I/\pi)\int_{-\infty}^{+\infty}d\sigma_R\, \Phi(\sigma)=\int_0^\infty dE\,r(E).\la{su}\ee
For the quantity in the right--hand side the usual sum rule is valid. Namely, taking into account Eq.~(\re{de}) one notices that
the right--hand side of Eq.~(\re{5}) is the square of the coefficient in the expansion of $\rho\chi_0(\rho)$ over $\chi_E(\rho)$.
Therefore, one has
\be \int_0^\infty dE\, r(E)=\int_0^\infty d\rho\,\rho^2 \chi_0^2(\rho)\equiv\langle\rho^2\rangle.\la{sr}\ee
In the $\sigma_I=5$ MeV case, the integration in the left--hand side of Eq.~(\re{su}) was performed in the range
between -40~MeV and 60~MeV with the step 0.5 fm and the integrals  from the asymptotic expressions 
\mbox{$\Phi(\sigma_R,\sigma_I)\simeq\langle\rho^2\rangle/\sigma_R^2$}
 over the intervals beyond this range were added to the
result. This gives for the left--hand side of Eq.~(\re{su}) the value of 10.15~fm$^2$ while the $\langle\rho^2\rangle$ value
is equal to 10.13~fm$^2$.

In Table 2 the trend of
convergence of the transform obtained 
is shown at some $\sigma_R$ values for \mbox{$\sigma_I=5$~MeV}.  
The quantity~$N_{max}$ denotes the number of the oscillator functions (\re{osc})
retained at solving Eq.~(\re{6}) while
their number 
retained in the expansions of the bound state entering its right--hand side was $N_{max}+1$ in all the cases. 

\begin{table}[h]
\caption{Dependence of the $\Phi(\sigma_R,\sigma_I)$ [fm$^2$ MeV$^{-2}$] values at $\sigma_I=5$ MeV on the
number $N_{max}$ of the functions (\re{osc}) retained in the calculation.}
\vspace{0.2cm}
\begin{tabular}{|c|c|c|c|}
\hline
$N_{max}$& $\sigma_R=-10$ MeV & $\sigma_R=8$ MeV&$\sigma_R=30$ MeV \\
\hline
24 & 2.9935134728858 $\cdot 10^{-2}$  & 0.309122264 &2.64204 $\cdot 10^{-2}$ \\
\hline
149& 2.9939341199509 $\cdot 10^{-2}$ & 0.309082197&2.62388 $\cdot 10^{-2}$  \\
\hline
299& 2.9939341199512 $\cdot 10^{-2}$  & 0.309082192  & 2.62383 $\cdot 10^{-2}$  \\
\hline
\end{tabular}
\end{table} 
At the $\sigma_R=8$~MeV value  in the Table the transform $\Phi$ reaches its maximum, up to the  
grid step in $\sigma_R$ which is equal to 0.5~MeV in the present case. The maximum position obtained is at variance with that 
in Ref.~\cite{suz} where, according to Fig. 7 there, the  $\sigma_R$ value at which $\Phi$ reaches its maximum
exceeds 10 MeV. 

We seek for the continuum spectrum wave function of Eq.~(\re{22}) in the form
\be  
{\bar \chi}_E(\rho)=
\sqrt{\rho}J_3(k\rho)+c_0\left[1-\exp\left[-(\rho/\rho_{cut})^2\right]\right]^3
\sqrt{\rho}N_3(k\rho)+\sum_{n=1}^{N_{max}}c_n\phi_n(\rho)\la{csp} ,
\ee
where \mbox{${\bar \chi}_E(\rho)=\left[(m/\hbar^2)^{1/2}\cos\delta\right]^{-1}\chi_E(\rho)$}
and $\phi_n$ are the functions~(\re{osc}).
The  $c_0$ and $c_n$ coefficients are to be found. One has \mbox{$c_0=-\tan\delta$}. The regularization
factor $[\ldots]$ in front of $\sqrt{\rho}N_3(k\rho)$
leads to the correct behavior~$\propto\rho^{l_1+1}$  of the corresponding term
at $\rho$ tending to zero. The $\rho_{cut}$ parameter has been taken to be 2.0~fm. Usually, the 
coefficients of such type expansions are obtained with the help of the Hulth\'en--Kohn type equations. Calculating matrix elements
that enter such equations encounters 
difficulties in the many--body case. Because of this, an alternative set of equations has been suggested~\cite{ze71}. 
We shall employ the latter set of equations here which will also provide a test of the approach.  In the present case,
these equations are projections of the Schr\"odinger equation for ${\bar \chi}_E(\rho)$ onto the set of oscillator 
functions~(\re{osc}) with \mbox{$n=1,\ldots,N_{max},N_{max}+1$}.

In Table 3  the trend of
convergence of the phase shift~$\delta$ 
thus obtained is shown at several energies. The notation~$N_{max}$ is as 
in Eq.~(\re{csp}).

\begin{table}[h]
\caption{Dependence of the phase shift $\delta(E)$ [deg]  on the
number $N_{max}$ of the functions~(\re{osc}) retained in the calculation.}
\vspace{0.2cm}
\begin{tabular}{|c|c|c|c|}
\hline
$N_{max}$& $E=1$ MeV & $E=10$ MeV&$E=20$ MeV \\
\hline
25 & 9.8070390 $\cdot 10^{-2}$& 68.9283508  &96.528079  \\
\hline
150& 9.9603267 $\cdot 10^{-2}$ & 68.9249288 &96.523602   \\
\hline
250& 9.9603273 $\cdot 10^{-2}$ & 68.9249289 & 96.523603  \\
\hline
\end{tabular}
\end{table} 

In Table 4 
the trend of convergence of the response function~$r(E)$ calculated directly according to Eq.~(\re{5})
 is shown at several energies. The notation~$N_{max}$ is as 
in Eq.~(\re{csp}) 
 while
the number of the oscillator functions~(\re{osc})
retained in the expansion of the bound state equaled~$N_{max}+1$ in all the cases. 

\begin{table}[h]
\caption{Dependence of the response function $r(E)$  [fm$^2$ MeV$^{-1}$] on the
number $N_{max}$ of the functions~(\re{osc}) retained in the calculation.}
\vspace{0.2cm}
\begin{tabular}{|c|c|c|c|}
\hline
$N_{max}$& $E=1$ MeV & $E=8$ MeV&$E=20$ MeV \\
\hline
25 & 4.434104 $\cdot 10^{-2}$& 1.3213&4.833 $\cdot 10^{-2}$  \\
\hline
250& 4.434102 $\cdot 10^{-2}$& 1.3210  & 4.834 $\cdot 10^{-2}$  \\
\hline
\end{tabular}
\end{table} 

The obtained response function and the quantities \mbox{$(\sigma_I/\pi)\Phi(\sigma)$} at $\sigma_I=2.5$ and 5~MeV 
are shown in Figure 1.
\begin{figure}[ht]
\centerline{\includegraphics*[scale=0.5]{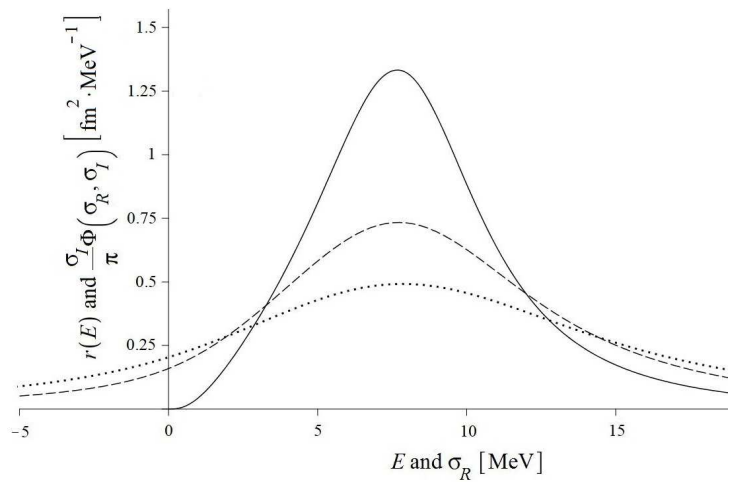}}
\caption{Full line: the response function $r(E)$. Dashed line: its integral transform $\Phi(\sigma_R,\sigma_I)$ at $\sigma_I=2.5$
MeV multiplied by $\sigma_I/\pi$. Dotted line: the same for $\sigma_I=5$ MeV.}
\end{figure}

\section{Inversion of the LIT}

When solving Eq.~(\re{ieq}) we, as usual, seek for the response function  in the form of an expansion over a set of basis 
functions~$f_n(E;\alpha)$ where $\alpha$ is a fall--off parameter, 
\be r(E)=\sum_{n=1}^{N}C_nf_n(E;\alpha).\la{exp}\ee
Substitution of this expression in the right--hand side
of Eq.~(\re{ieq}) gives the trial transform~$\Phi_{tr}$,
\ber \Phi_{tr}(\sigma_R,\sigma_I)=\sum_{n=1}^{N}C_n{\bar f}_n(\sigma_R,\sigma_I;\alpha),\la{ptr}\\
{\bar f}_n(\sigma_R,\sigma_I;\alpha)=\int_0^\infty dE\,\frac{f_n(E;\alpha)}{(E-\sigma_R)^2+\sigma_I^2}.\la{int}\eer
One imposes the minimum condition 
\be \sum_i[\Phi((\sigma_R)_i,\sigma_I)-\Phi_{tr}((\sigma_R)_i,\sigma_I)]^2W(i)={\rm min}.\la{min}\ee
The minimization is performed with respect to the  $C_n$ and $\alpha$ parameters entering $\Phi_{tr}$
on a sufficiently
dense grid   $(\sigma_R)_i$ in some range of $\sigma_R$ values.
The quantity $W(i)$ is a weight function. The condition (\re{min})
leads to the set of linear equations for the $C_n$ parameters. 

In the exact arithmetic the inversion results would not depend on the 
choice of the mentioned   range of $\sigma_R$ values since both sides
of  Eq.~(\re{ieq}) are analytic functions of $\sigma_R$. However, in practice this choice matters. 
In Ref. \cite{suz} it was recommended to employ a very wide range of these values, such that provides the fulfillment of
the relation  (\re{su}). We believe that such a choice is not an optimal one. Indeed, at such a choice
much weight is given to the large $|\sigma_R|$ wings of $\Phi(\sigma_R,\sigma_I)$. But, in accordance with Eqs.~(\re{ieq}) and 
(\re{sr}), $\Phi(\sigma_R,\sigma_I)$ behaves at these wings as $\langle\rho^2\rangle/(\sigma_R)^2$, i.e. in a universal
way, and thus does not 
provide substantial information on the  behavior of $r(E)$. 

The choice of the  range of $\sigma_R$ values  for the inversion purposes is to be related with the interval  $0\le E\le E_{max}$
of $E$ values on
which we want to get the $r(E)$ response.
Below we employ the \mbox{$-2\sigma_I\le\sigma_R\le E_{max}+2\sigma_I$} range to this aim.

In Ref.~\cite{suz} the following basis set has been used,
\be f_n(E;\alpha)=E^3e^{-\alpha E/n}.\la{exp1}\ee
Below we shall perform the inversion in two versions. In one of them, we shall use exactly the same set~(\re{exp1}).
In the other version, we shall modify this set which will lead to simplifications.

As to the choice of the set~(\re{exp1}), one may note that inversion 
is frequently facilitated by incorporating the true low--energy
behavior of the response 
into the basis functions.  In the case of two fragments above the threshold
with no Coulomb inter--fragment interaction this behavior is $E^{l_1+1/2}$ which is
seen from Eq.~(\re{33}).
 The $E^3$ factor in Eq.~(\re{exp1}) reproduces this behavior.
  The  set of basis exponentials from Eq.~(\re{exp1}) is complete in the sense of both  $L^2(0,\infty)$
and $C(0,\infty)$ norms.  This follows from the M\"unz theorem, see e.g. \cite{Akh}.

The accuracy of inversion should increase as the number $N$ of basis functions in the expansion (\re{exp}) increases.
As in the previous work, see e.g.~\cite{rev}, stability of the results of the inversion in some range of $N$ values  serves
as a criterion of its reliability.
However, as it is known,  
stability arising with an increase of $N$ 
may  well be violated at its further increase due to the fact that not an exact transform but an approximate one is fitted.
Besides, with the basis functions of Eq.~(\re{exp1}) it is impossible to perform calculations at too large $N$ values because of
round--off errors, see below.

We shall study the  $\sigma_I=5$~MeV and  $\sigma_I=2.5$~MeV cases. The first of these cases was 
considered in Ref.~\cite{suz}. Such choices of the $\sigma_I$ 
values would be reasonable when one deals with responses having widths
comparable with that of the present $r(E)$ response. Use of smaller $\sigma_I$ values in the many--body case
would require more effort in order to 
solve the inhomogenous equation like Eq.~(\re{6}) with the same accuracy.

Inversion with the basis set~(\re{exp1}) occurs to be the most difficult in the region of small energy.
While in Ref. \cite{suz} the weight function $W(i)$ entering the fitting procedure of Eq.~(\re{min}) was taken to be unity, 
here,  to
improve the inversion in the mentioned  region in the case of set~(\re{exp1}),  
 $W(i)$ 
 has been 
chosen  
as follows. Below the point $\sigma_R=\sigma_0$ of the maximum, at a given~$\sigma_I$, of
$\Phi(\sigma_R,\sigma_I)$  the weight function $W(i)$ was taken to be
\mbox{$[\Phi(\sigma_0,\sigma_I)/\Phi((\sigma_R)_i,\sigma_I)]^2$} and it was taken to be unity beyond this point.

Even at this choice of $W(i)$ it is necessary to retain rather many basis functions (\re{exp1}) in Eq.~(\re{exp}) 
to get a good inversion at 
small energy. It occurs that 
when the number of basis functions increases the expansion coefficients become very large
in magnitude. In the present case they reached the values about $10^{10}$ at the highest $N$ values we employed.
Corresponding contributions strongly cancel each other.
This feature has not been noticed so far. Because of it, 
the inversion  was performed via calculations with the quadrupole precision in this version. 
Also the integrals~(\re{int}) are to be calculated here with 
high accuracy. They were expressed~\cite{nir} in terms of the incomplete gamma function of a complex argument.
However, the accuracy with which this function is provided by existing codes is not known. We  calculated these integrals
numerically in the intervals \mbox{$0\le E\le E_{max}$} at \mbox{$E_{max}=130n/\alpha$ MeV} with the relative 
accuracy of~$10^{-21}$.
All this  provided  sufficient stability of the final results at $E\ge3$~MeV. However, at lower energies the stability 
remains incomplete 
irrespective to accuracy of the integration. At $\sigma_I=5$~MeV round--off errors may influence the result
in the first non--zero decimal 
place at $E=1$~MeV and in the second non--zero decimal 
place at \mbox{$E=2$ MeV}. At $\sigma_I=2.5$~MeV they may influence the result in the second non--zero decimal 
place at $E=1$~MeV and in the third non--zero decimal 
place at $E=2$~MeV.

Provided that the overlap integrals of basis functions are known exactly, as in the present case,
it is probably 
possible to get rid of the large  $|C_n|$ values using a  basis set $f_n$ in Eq.~(\re{exp}) which is orthonormalized.
In this case, the $C_n$ coefficients for the corresponding expansion of the exact~$r(E)$ are such that 
the sum $\sum_{n=1}^N C_n^2$ 
is bounded from above by the integral from $r^2(E)$ over all
the energies. Therefore, the $|C_n|$ values cannot be large. Probably this refers also to the approximate $C_n$
coefficients determined from the fit.
However, this version have not been tried.

In the above version of the calculation, 
the optimal value of the $\alpha$ parameter entering functions~(\re{exp1}) was searched first on a grid. After that, the 
minimum of the expression (\re{min}) was looked for on a smaller $\alpha$ interval. However, at large
$N$ values the arising dependence of the quantity (\re{min}) on $\alpha$ becomes a fluctuating one due to a
strong cancellation between $\Phi$ and $\Phi_{tr}$. 
 Because of this, it is not possible to find the absolute minimum. 
Anyway, the obtained fits  to $\Phi(\sigma_R,\sigma_I)$ are very precise.

\begin{table}[h]
\caption{The exact response $r(E)$ and the responses obtained via inverting the LIT of $r(E)$ 
at $\sigma_I=5$~MeV with various numbers
$N$ of basis functions  retained in the expansion (\re{exp}). 
The energy $E$ is in~MeV.}
\vspace{0.2cm}
\begin{tabular}{|c|c|c|c|c|c|c|}
\hline
&&\multicolumn{5}{|c|}{$N$}\\
\hline
$E$ & exact & 30 &35 &40 &45 &50\\
\hline
  1&     0.044  &    -0.042   &    0.012 &      0.015  &     0.009  &     0.005\\
   2 &    0.177  &     0.215   &    0.178  &     0.196  &     0.198   &    0.237\\
  3 &     0.354  &     0.361   &    0.363  &     0.357  &     0.356  &     0.349\\
  4 &     0.564  &     0.559   &    0.559  &     0.561  &     0.562  &     0.564\\
   8&    1.321 &      1.322     &  1.321   &    1.321  &     1.321 &      1.322\\
   12&    0.452 &      0.452  &     0.452  &     0.452  &     0.452  &     0.452\\
   16&    0.130 &      0.130  &     0.130   &    0.130   &    0.130  &     0.130\\
    20&   0.048  &     0.048   &    0.048  &     0.048   &    0.048   &    0.048\\
\hline
\end{tabular}
\end{table}

In Table V the results of the above described
inversions at $\sigma_I=5$~MeV are presented at some  energies~$E$
and various numbers~$N$
of basis functions retained in the expansion (\re{exp}). The exact response~$r(E)$ is presented in the second
column and   
the responses obtained from the inversion of the LIT are shown in columns from three to seven. 
Note that at  $E\ge4$~MeV at least
the two--digit  accuracy of inversion has been obtained also for all the $E$ values 
not shown in the Table  in the range $E\le20$~MeV considered.
As to lower energy, the results are also rather accurate at $E=3$~MeV, and at $E=2$~MeV 
they are of a
moderate accuracy. 
These results are different from those of Ref.~\cite{suz} where large or substantial deviations from
the true response were found at all the energies.

In Table VI the corresponding results at $\sigma_I=2.5$~MeV are presented. The results  in 
the range considered for energies not presented in the table
are quite similar. Here we show also the results of the inversions at smaller
$N$ values which
deviate from those in the region of stability with respect to $N$. In this case, stability is reached at $N$
values smaller than in the preceding $\sigma_I=5$~MeV case. It is also seen 
that in the present case stability takes place also at small energies. 
This is in line with the fact that   the resolution
of the Lorentz kernel here is higher than in the preceding case.

\begin{table}[h]
\caption{The exact response $r(E)$ and the responses obtained via inverting the LIT of $r(E)$  
at $\sigma_I=2.5$ MeV. The notation is as in Table V.}
\vspace{0.2cm}
\begin{tabular}{|c|c|c|c|c|c|c|c|c|}
\hline
&&\multicolumn{7}{|c|}{$N$}\\
\hline
$E$ & exact & 10 &15 &20 &23 &25&27&28\\
\hline
 1&      0.044 &      0.107  &     0.113  &     0.029  &     0.044  &     0.044   &    0.041  &     0.043\\
  2&     0.177  &     0.118   &    0.158 &      0.167  &     0.186  &     0.183  &     0.176  &     0.184\\
  3&     0.354 &      0.390   &    0.358 &      0.362  &     0.355  &     0.355  &     0.356  &     0.355\\
  4&     0.564 &      0.571 &      0.561 &      0.560 &      0.563 &      0.563 &      0.563  &     0.563\\
 8&      1.321  &     1.311  &     1.321   &    1.321   &    1.321   &    1.321   &    1.321  &     1.321\\
 12&      0.452    &   0.454    &   0.452  &     0.452  &     0.452   &    0.452   &    0.452   &    0.452\\
 16&      0.130  &     0.131    &   0.130  &     0.131   &    0.131  &     0.131    &   0.130 &      0.131\\
 20&      0.048&       0.048 &      0.049  &     0.049   &    0.049 &      0.049   &    0.048   &    0.049\\
\hline
\end{tabular}
\end{table}

It should be noted that the above precise calculations were feasible solely due to the fact that
the input transform was known with a high precision. In the usual many--body applications
one works with a transform that is 
considerably less precise than in the present case.
In what follows a simpler version of the inversion is presented. 
As we shall see, in this case so high a precision of the calculation is not required.
Accordingly,
for the inversions that follow we use an upper integration limit of 100 MeV in Eq.~(17)
(tiny contributions beyond 100 MeV are neglected).
The weight function $W(i)$ is taken to be unity.
In addition a sowewhat different search for the fall-off parameter $\alpha$ is implemented. 
For a predefined set of $\alpha$ values the minimization of 
Eq.~(\ref{min}) is performed with respect to the linear parameters $C_n$.
For any number of basis functions $N$ the parameter set leading to the smallest value
of Eq.~(\ref{min}) is taken. Note that the response function is positive definite, therefore
we exclude parameter sets that lead to a negative response in the energy interval of interest.
The interval of $\sigma_R$ values we consider here runs from zero up to $E_{max}+2\sigma_I$.

The search for an optimal $\alpha$ value is made as follows. We run $\alpha$ 
over a large grid of possible values and determine
the error given by the left-hand side of Eq.~(\ref{min}).   
In the present case the search was made with the following values:
\be
\alpha(j) = {\frac {1000}{j}}
\ee
with $j=1,2,...,1500$. We select the best fit among the
1500 trials as inversion result, which, in addition, has to fulfil the above defined positiveness condition. 

\begin{figure}[ht]
\centerline{\includegraphics*[scale=0.5]{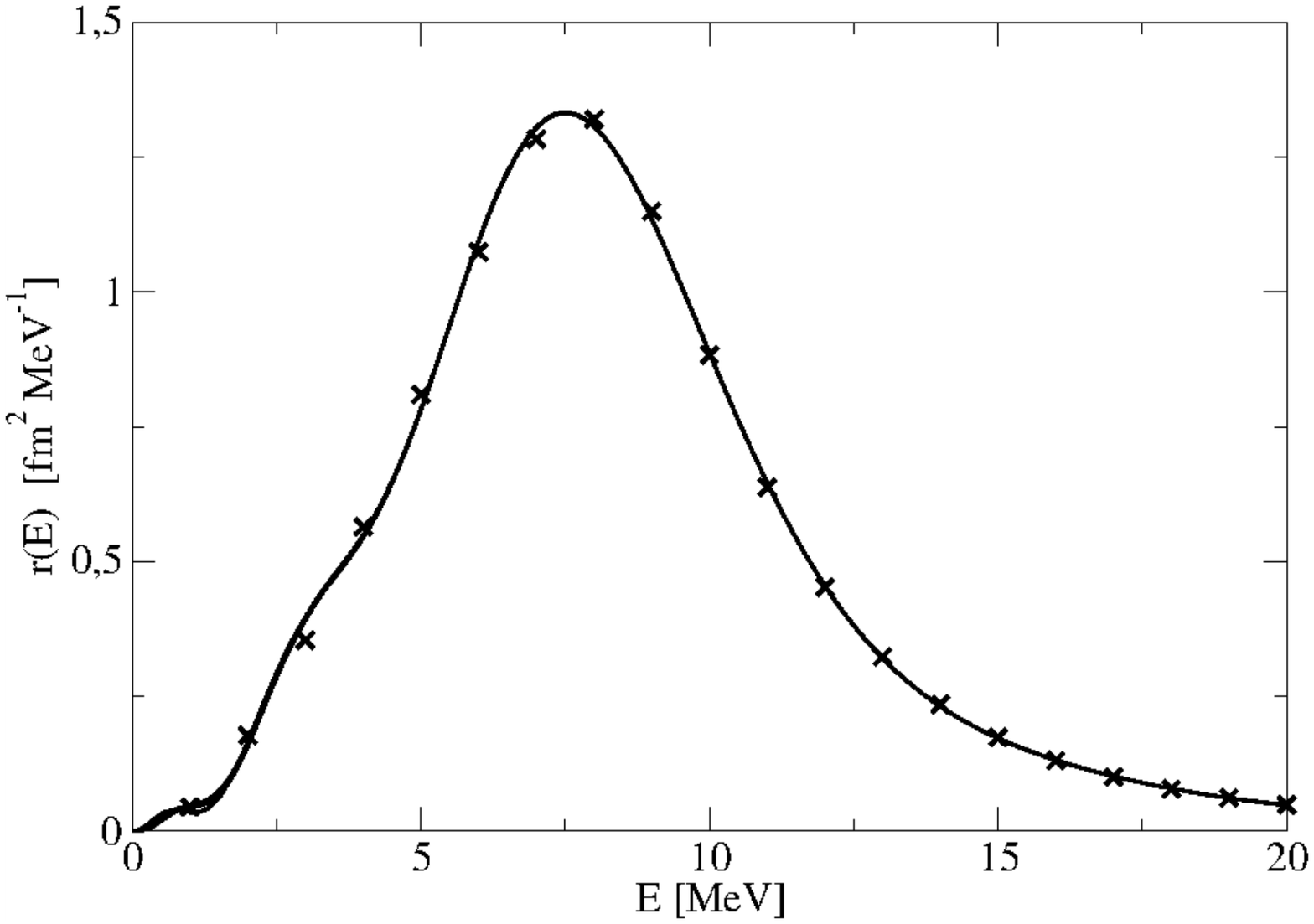}}
\caption{Inversion results for $N$ ranging from 10 to 20 ($\sigma_I=5$ MeV). Any of the eleven inversions
is represented by a solid line, true response is depicted by crosses (x).}
\end{figure}
After having determined the "best" response functions for the various number of basis functions
one compares the obtained results. In a perfect inversion of an analytically known
transform $\Phi$ the precision of the inversion improves with a growing number $N$ of basis
functions. In practise, due to numerical errors both in the calculation of $\Phi$
and in the inversion, one should observe a scenario already described above after Eq.~(\ref{exp1}): 
With an increase of $N$ one should find a rather stable inversion result for a limited range of $N$ values, then,
with a further increase of $N$ the stability is lost.\\
  
In Fig.~2 we show the inversion results for the $\sigma_I=5$ MeV case, where $N$ runs from 10 to 20.
One observes a rather stable result, in fact inversions with $15 \le N \le 20$ are almost
identical. Unsatisfying is the somewhat oscillatory behaviour below 5 MeV. In fact comparing with
the true response one finds differences up to the peak region. 

One can try to improve the inversion
using a smaller $\sigma_I$ value. Taking $\sigma_I=2.5$ MeV we obtain very stable inversion results with
$N=15,16,17$, whereas for even higher $N$ stronger low-energy oscillations set in. In Fig.~3 we
compare these results with the stable result obtained for $\sigma_I=5$ MeV. The comparison is made
only at lower energies, since at higher energies results are almost identical. For $\sigma_I=2.5$ MeV one notes 
a reduction of the oscillatory low-energy behaviour with an improvement of the result,
particularly visible at 3 and 5 MeV.
\begin{figure}[ht]
\centerline{\includegraphics*[scale=0.5]{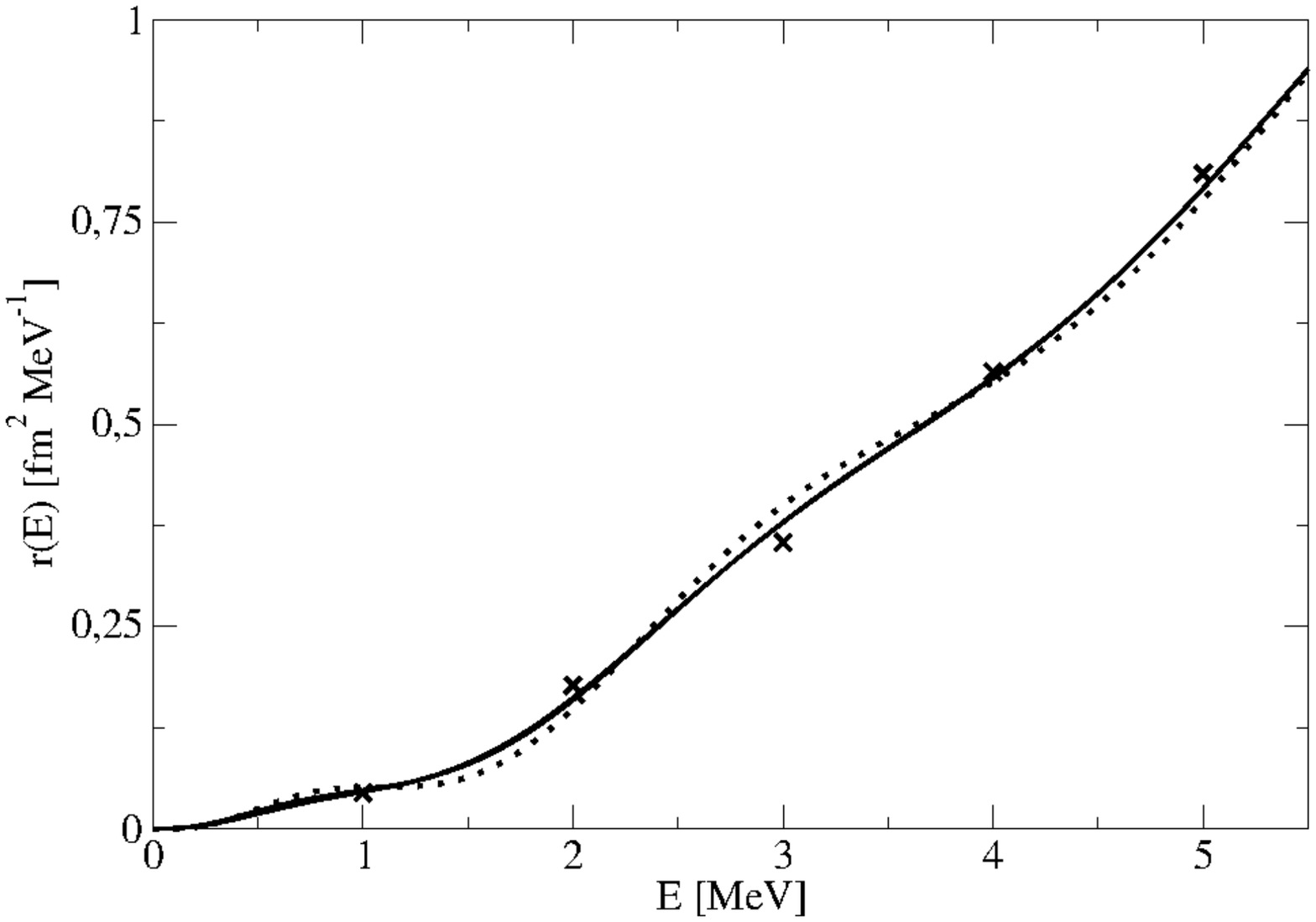}}
\caption{Comparison of inversion results with $\sigma_I=2.5$ MeV (solid line) and $\sigma_I=5$ MeV (dotted line).
Crosses (x) as in Fig.~2.}
\end{figure}

Thus, in order to further improve
the quality of the inversion one should work with a considerably smaller $\sigma_I$. In general,
it is not easily possible to calculate the transform $\Phi$ with a sufficient precison for a
much smaller $\sigma_I$. However, even in the present case a further improvement can be made.
As already stated the inversions show an oscillatory behaviour at lower energies. This
points in the direction that the chosen basis set is not very efficient. To illustrate this better
we show in Fig.~4 the inversion result for $\sigma_I=2.5$ MeV and $N=8$. One observes that there
is a maximum at about 0.6 MeV and a minimum at about 1.2 MeV. Such a structure of the response at
threshold is rather improbable and in fact for the above mentioned stable inversion result with
$N=15,16,17$ such a strong low-energy
oscillation has vanished. To avoid a fake strong low-energy rise one can take a different low-energy
behaviour of the basis functions used for the inversion.
In this connection one may note that in reality the $E^3$ behavior of the  response takes place only in a quite
narrow energy region close to zero. Deviations from this behavior are large already at e.g.~$E=0.2$~MeV.
(But if one drops all the oscillator basis functions
except the lowest one in the expansion obtained of the bound state wave function then
the $E^3$ behavior of the spectrum will take place in a rather wide range of energy. With the five percent accuracy it 
is then valid up to energies higher than $E=1$~MeV.) 
\begin{figure}[ht]
\centerline{\includegraphics*[scale=0.5]{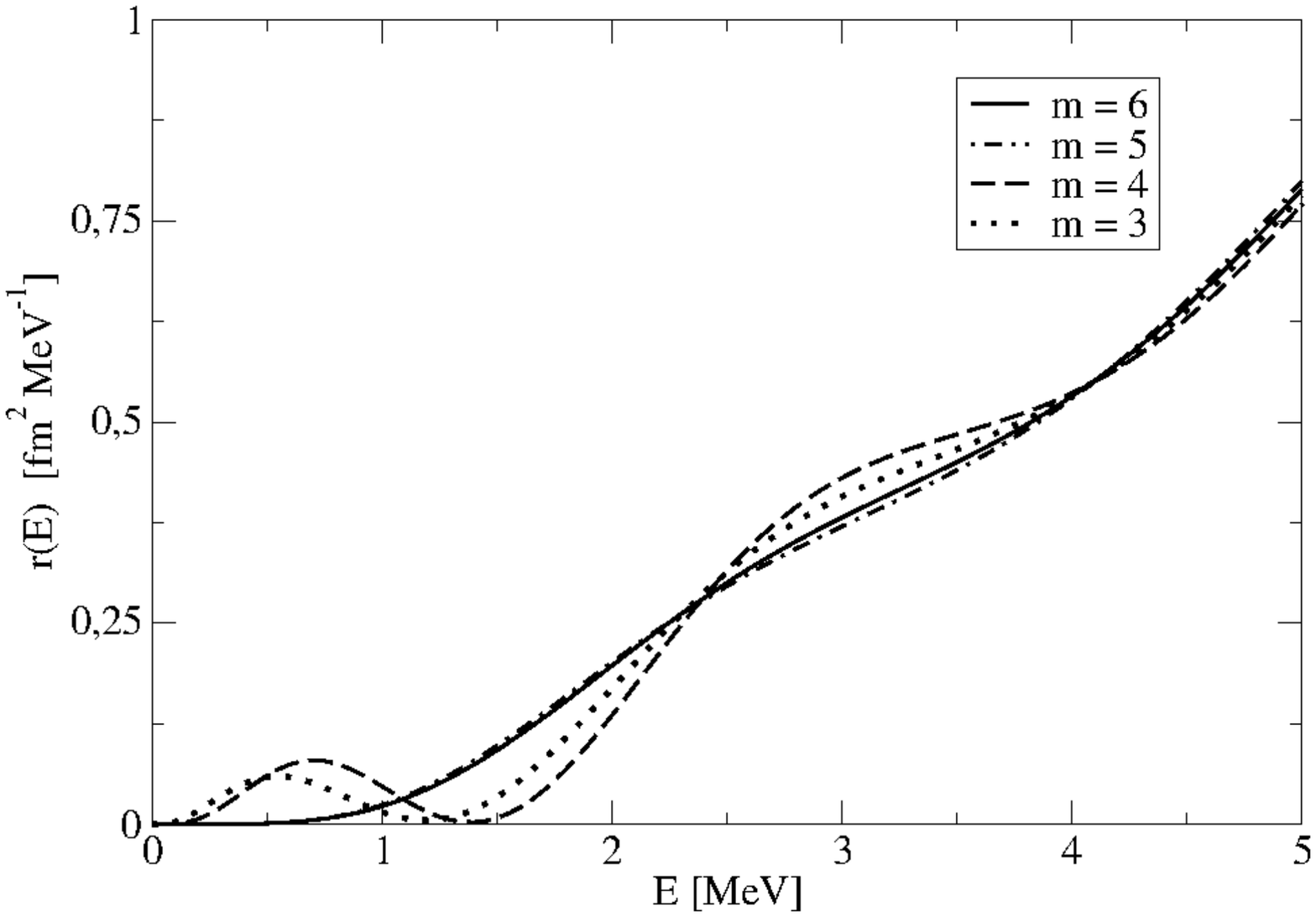}}
\caption{Inversions for $\sigma_I=2.5$ MeV and $N=8$ for various sets of basis functions, 
\mbox{$f_n(E;\alpha,m)=E^me^{-\alpha E/n}$}, with $m=3,4,5,6$.}
\end{figure}

In Fig.~4 we also show results where
the low-energy rise of $E^3$ of the $f_n$ of Eq.~(\ref{exp1}) is changed into $E^m$ with $m=4,5,6$.
One sees that the unwanted low-energy behaviour goes away with $m=5,6$. Studying better the case with 
$m=6$ we find a stable inversion with $N=14,15,16$. The corresponding results are shown in Fig.~5.
Now one observes a very nice agreement between true response and inversion result also at low energies.  
\begin{figure}[ht]
\centerline{\includegraphics*[scale=0.5]{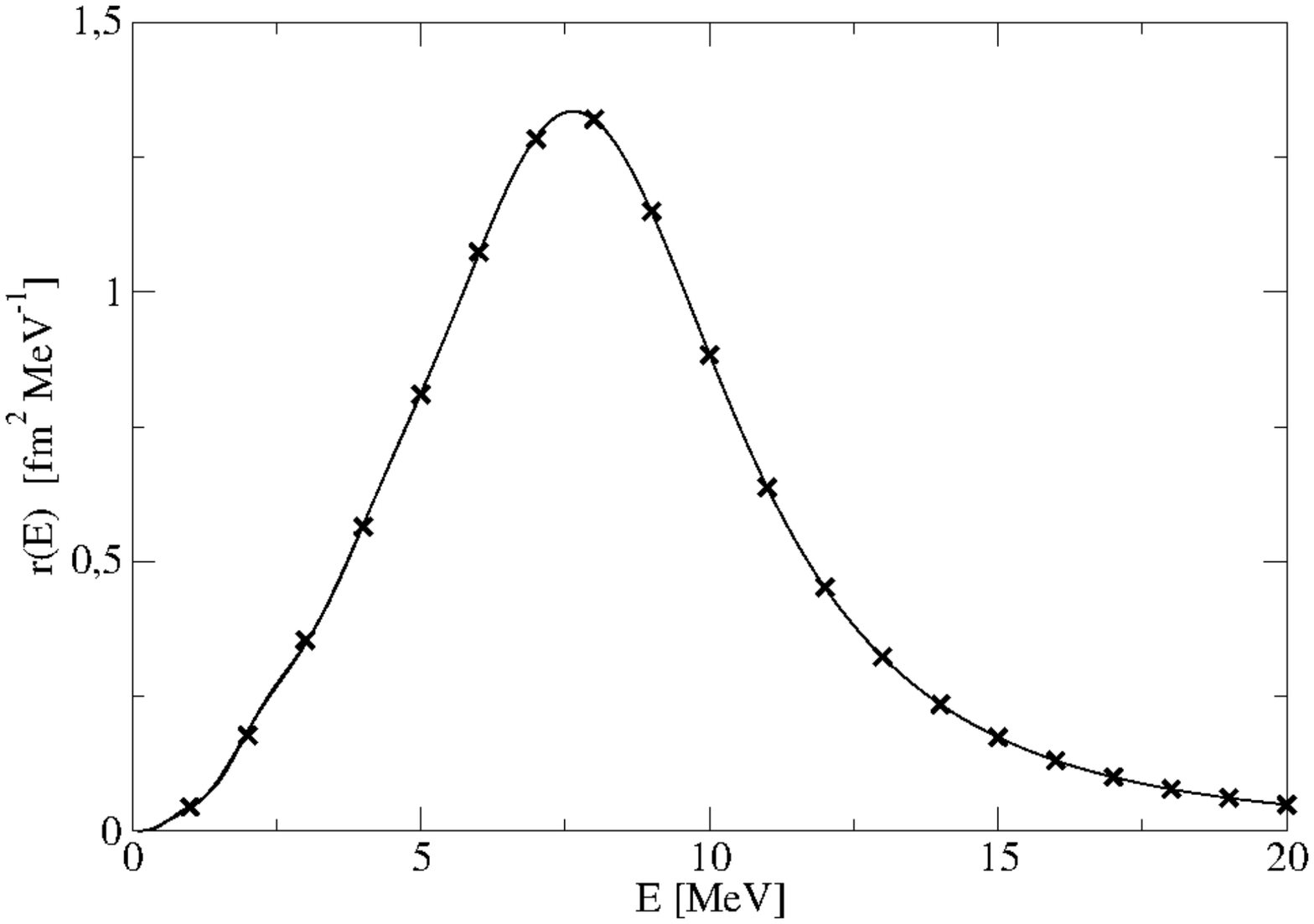}}
\caption{Inversions for $\sigma_I=2.5$ MeV and $N=14,15,16$ (solid lines) with basis function sets 
\mbox{$f_n(E;\alpha,m=6)$}. Crosses (x) as in Fig.~2.}
\end{figure}

\section{summary}

In Ref. \cite{suz} the question was addressed whether the response function provided by the model
of that work may be obtained with a resonable accuracy via inverting its LIT calculated  with
the bound--state type method.  The answer proved to be negative. In the present work, the problem was reconsidered
and, at variance with the results of Ref.~\cite{suz}, accurate approximations to the true response have been obtained
with that method.

Our calculation differs from that of Ref. \cite{suz} both in its input and in the inversion procedure. As to the input,
the LIT calculated from the inhomogeneous equation in the present work proves to be different from
that in Ref.  \cite{suz}. Also the ranges of values of the transform  employed for the inversion have been chosen
differently.

We started with the same inversion procedure as in Ref. \cite{suz} except for the weight function used 
at fitting the transform. In particular, the basis set used at performing the inversion was the same. 
In this way
we succeeded to obtain the responses 
of an acceptable accuracy.   Rather many basis functions were to be retained in order to reach
the inversion stability at low energy. 

However, transforms pertaining to few--body calculations normally have  considerably lower accuracy
than in the present case. To invert them, use of such an amount
of basis functions is not possible since the stability of inversion results would then be lost and 
 a non--physical oscillating function would be selected in the course of the inversion as the output
response.  

The next variant of our inversion procedure was the following.   The response function that provides
the best fit to an input transform is selected in the course of the inversion. We imposed the condition that 
 response functions taking non--physical
negative values are discarded at searching for the best fit. 
At this condition, the stability of the inversion results has been reached
at lower numbers of basis functions than above. The responses obtained still exhibit
 somewhat oscillatory behavior at low energy.

Finally, we have modified the basis set used for the inversion.  The inital basis set reproduced the low--energy behavior of the true
response. Such a condition was usually imposed in calculations in the literature and seemed to improve \cite{ELO99}
the inversion results. However, in the present case this proved to be not true and just because of this condition it
was necessary to retain a large amount of basis functions in the first above mentioned variant of the inversion
procedure. The peculiarity of the present problem is that the asymptotic low--energy behavior of the response takes place
only in a very limited energy range. Therefore, we have modified the factor entering the basis functions which describes
the low--energy behavior of the response sought for. We have chosen this factor from the condition that non--physical 
oscillations of the response are excluded already at small number of basis functions used for the inversion. This
improved our outcome at low energy leading to a nice agreement between the inversion result and the true response.

Acknowledgement of support is given to
RFBR Grant No. 18--02--00778 (V.D.E and V.Yu.S.).

\end{document}